*PRE-PRINT*

# Graphing Methods for Kendall's τ

Nicholas D. Edwards[a]; Enzo de Jong[a]; Stephen T. Ferguson[a]*

[a]*Department of Biological Sciences, Vanderbilt University, Nashville, TN 37235, USA.*

*Correspondence to: stephen.t.ferguson@vanderbilt.edu

# Graphing Methods for Kendall's τ

**Abstract**: Ranked data is commonly used in research across many fields of study including medicine, biology, psychology, and economics. One common statistic used for analyzing ranked data is Kendall's τ coefficient, a non-parametric measure of rank correlation which describes the strength of the association between two monotonic continuous or ordinal variables. While the mathematics involved in calculating Kendall's τ is well-established, there are relatively few graphing methods available to visualize the results. Here, we describe a visualization method and provide an interactive app for graphing Kendall's τ which uses a series of rigid Euclidean transformations along a Cartesian plane to map rank pairs into discrete quadrants. The resulting graph provides a visualization of rank correlation which helps display the proportion of concordant and discordant pairs. Moreover, this method highlights other key features of the data which are not represented by Kendall's τ alone but may nevertheless be meaningful, such as the relationship between discrete pairs of observations. We demonstrate the effectiveness of our approach through several examples and compare our results to other visualization methods.

**Main Text**

Ranked data is a fundamental statistical data type that is widely used in diverse fields of research, including medicine (Weber and Titman 2019), biology (Ferguson et al. 2023), psychology (Fincham et al. 2023), and economics (Pirttilä and Uusitalo 2010). This type of data provides information on the order and relative position of each observation in a dataset, allowing researchers to analyze the rank relationship between variables and make comparisons between groups. Due to its prevalence in research, various statistical methods have been developed to analyze ranked data, making it an essential aspect of statistical analysis (Friedman 1937, 1939; Kendall 1938; Milton 1940; Spearman 1987, Wilcoxon 1945). Rank correlation coefficients are one such method and refer to statistical analyses that measure the similarity between two sets of ranked data examining a given variable. Here, we focus our discussion on Kendall's τ (Kendall 1938, 1955), a widely used non-parametric rank correlation coefficient for which relatively few visualization methods have been described (Holmes 1928; Wegman 1990; Davis and Chen 2007).

Kendall's τ is a measure of rank correlation that describes the strength of the association between two monotonic continuous or ordinal variables by calculating the proportion of concordance and discordance among pairwise ranks (Kendall 1938, 1955). Here, for a given set of ranks $(x_1, y_1), \ldots, (x_n, y_n)$, a pair of observations $(x_i, y_i)$ and $(x_j, y_j)$ is considered concordant if $sgn(x_j - x_i) = sgn(y_j - y_i)$. Alternatively, if $sgn(x_j - x_i) = -sgn(y_j - y_i)$, then the pair is considered discordant. Tie values arise when either sign function is equal to 0. The total number of pairwise comparisons that can be made is therefore: $n = c + d + t_x + t_y + t_{x,y}$ where $c$ is the total number of concordant pairs, $d$ is the total number of discordant pairs, $t_x$ is the total number of pairs where only $sgn(x_j - x_i) = 0$, $t_y$ is the total number of pairs where only $sgn(y_j - y_i) = 0$, and $t_{x,y}$ is the total number of pairs where both sign functions equal 0. Thus, Kendall's τ can be expressed as follows: $\tau = \frac{c-d}{\sqrt{(c+d+t_x)*(c+d+t_y)}}$. The resultant τ value can range from 1 to -1 where 1 indicates a perfect concordant relationship, -1 indicates a perfect discordant relationship, and 0 indicates no relationship between the ranks.

Despite being a fundamental and broadly used measure of rank correlation, relatively few methods have been described for visualizing and graphically representing Kendall's τ (Holmes 1928, Wegman 1990, Davis and Chen 2007). This limitation can make it difficult to identify patterns or trends in the data. Interestingly, the earliest method for visualizing the results of Kendall's τ utilized parallel coordinates and was published in 1928 (Holmes 1928), predating the formal description of the statistic which was put forth by Kendall ten years later in 1938 (Kendall 1938). More recent methods have relied on the use of a Cartesian plane (Davis and Chen 2007). In this system, coordinates corresponding to the values of each rank set are plotted along the x and y axis, respectively. Lines are then drawn connecting all possible pairwise combinations of coordinates, and a color-code is applied based on the slope of each line such that concordant pairs will have a positive slope, discordant pairs will have a negative slope, and tie values will be either horizontal or vertical (Davis and Chen 2007). One potential limitation of these existing methods, however, is that large data sets result in a high degree of overlap between the lines drawn on the parallel coordinates or the Cartesian plane. This may obscure important details or otherwise create visual clutter which may be difficult to interpret. Here, we build on these methods by presenting several alternative visualization strategies for graphing Kendall's $\tau$ which are appropriate for both small and large data sets and which can be combined to highlight different features of the data. In addition to the previously described visualization methods discussed above (Holmes 1928, Davis and Chen 2007), we hope to provide a supplementary cache of graphical options which may be particularly meaningful when a researcher wants to glean a deeper understanding of the structure and organization of the data which underlies the quantitative $\tau$ value statistic.

*Graphical Transformation Procedures*

In order to visually represent Kendall's $\tau$, we used a series of rigid Euclidean transformations along a Cartesian plane. To illustrate this process, consider two hypothetical sets of ranked data where $x_0 = [20, 86, 35, 55, 60, 85, 8, 15]$ and $y_0 = [40, 78, 80, 35, 25, 15, 19, 93]$ which can be represented as coordinates along the x- and y-axis, respectively (Fig. 1A; Supplementary File 1). Lines can then be drawn connecting all possible pairwise combinations of coordinates which correspond to the comparisons made when evaluating Kendall's $\tau$ (Fig. 1A). Each line is then translated such that the coordinate with the lowest x-value—or the coordinate with the lowest y-value in the event of a tie—is relocated to the origin, and then rotated by $n$ degrees such that $n = 2 * \theta$ (Fig. 1B, Supplementary File 2). As a result of this rigid transformation, various features of the data underlying Kendall's $\tau$ become spatially organized along the graph: all lines corresponding to concordant pairs terminate above the x-axis in Quadrant I or Quadrant II; all lines corresponding to discordant pairs terminate below the x-axis in Quadrant III or Quadrant IV; and all tie values terminate along the x-axis (Fig. 1B). Note that deciding which coordinate to relocate to the origin, and whether or not to rotate these lines, is arbitrary. As a result, one can reasonably create different graphical representations of the same data using translation alone which provides greater flexibility when graphing data (Fig. S1A-B).

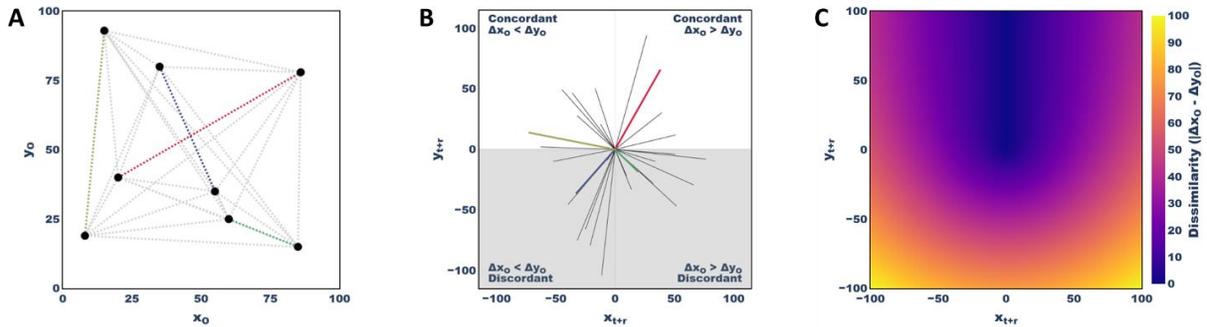

Figure 1. Method for visualizing Kendall's $\tau$ using rigid Euclidean transformations. **A.** Representative ranked data plotted as coordinates on a Cartesian plane. Each coordinate represents an item which has been ranked by two entities—$x_0$ or $y_0$—and the lines between coordinates signify the pairwise comparisons made during the evaluation of Kendall's $\tau$. **B.** The result of translating pairs of observations to the origin and then rotating the lines by doubling their angle ($2*\theta$); note that the axes labels have been updated to denote this transformation ($x_{t+r}$ or $y_{t+r}$). The area above the x-axis in white denotes concordant pairs, and the area below the x-axis in gray represents discordant pairs. **C.** A heatmap demonstrating how dissimilarity within and between rankings, as measured by the absolute difference between $\Delta x_0$ or $\Delta y_0$, changes with respect to location along the Cartesian plane.

Another unique feature of these representations is that the absolute difference in the original, untransformed rank values between any given pair of observations, which we have termed dissimilarity, is reflected by the spatial position of the terminal endpoint and its distance from the origin (Fig. 1C, Fig. S1C-D). For example, consider two different sets of untransformed observations: 1. $x_1 = [1, 2]$ and $y_1 = [2, 1]$; and 2. $x_2 = [1, 50]$ and $y_2 = [50, 1]$. While both rank sets contain a discordant pair, there is a difference in the relative magnitude of the ranks. In the first example, the dissimilarity between $x_1$ and $y_1$ is 2 ($|\Delta x_1 - \Delta y_1| = |(1 - 2) - (2 - 1)| = 2$). This difference is relatively small compared to $x_2$ and $y_2$, where the dissimilarity is 98 ($|\Delta x_2 - \Delta y_2| = |(1 - 50) - (50 - 1)| = 98$). Notably, these features map to discrete quadrants of the graph (Fig. 1B-C, Fig. S1). When translating and rotating observations, if $\Delta x_O > \Delta y_O$, then the endpoint of the line will terminate on the right side of the y-axis in Quadrant I for concordant pairs or Quadrant IV for discordant pairs (Fig. 1B). When $\Delta x_O < \Delta y_O$, the endpoint of the line will terminate on the left side of the y-axis in Quadrant II for concordant pairs or Quadrant III for discordant pairs (Fig. 1B). While this information does not directly influence the calculation of Kendall's τ, it can be especially meaningful when researchers want to examine the data underlying discrete comparisons. Ultimately, what constitutes a meaningful difference between ranks and the relevance of this information depends on the particular questions being asked by the researcher. We illustrate one such circumstance using real-world data below.

## *Real-World Example (Military vs. R&D Expenditures)*

To better illustrate this graphing method and its interpretation, we examined real-world ranked data using governments' expenditures on the military vs. research and development (R&D) in 2020 reported as a percentage of gross domestic product (GDP; Supplementary File 1) (World Bank N.d.a, N.d.b). When examining a graphical representation of this data, and even prior to calculating the formal τ value, it is apparent that there are an approximately equal number of concordant pairs (i.e., lines above the x-axis) and discordant pairs (i.e., lines below the x-axis) with a slight skew towards discordance (Fig. 2A). Indeed, the τ value for this dataset is -0.02, suggesting an absence of association between the rank order of military and R&D expenditures across countries. Beyond this fundamental representation of Kendall's τ, it is possible to extend this analysis by examining discrete concordant or discordant comparisons which are associated with the lines connecting coordinates on the graph. Consider, for example, the line labelled B on the graph (Fig. 2A) which corresponds to the data represented in Fig. 2B. This line, along with many nearby neighbors, represent pairs of countries in which both military and R&D expenditure were higher in one country compared to the other (i.e., concordance) and the magnitude difference in spending for each category was about the same (i.e., low dissimilarity). Here, Israel invested a considerably greater percentage of their GDP on both military and R&D compared to Moldova (Fig. 2A-B). In contrast, at line C (Fig. 2A), one country invested more heavily in military expenditures while the other invested more heavily in R&D (i.e., high dissimilarity) (Fig. 2C). Oman spent a much higher percentage of their GDP on the military while the Republic of Korea invested more heavily in R&D (Fig. 2C). At line D (Fig. 2A), R&D expenditure was approximately equal for both countries, but one country (Oman) invested considerably more in military expenditures compared to the other (Mauritius) (Fig. 2D). At line E (Fig. 2A), the opposite trend is observed. Military expenditure was approximately equal for both countries, but one country (Israel) invested more heavily in R&D compared to the other (Azerbaijan) (Fig. 2E).

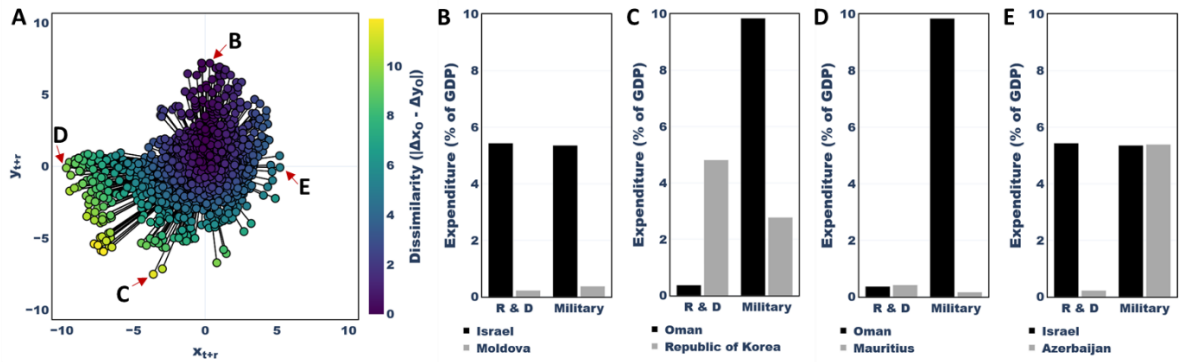

Figure 2. A representative, real-world example using governments' expenditures on military vs. R&D. **A.** A visual representation of Kendall's τ using military and R&D expenditure (as a percentage of GDP) from 62 different countries (Supplementary File 1) (World Bank, N.d.a, N.d.b). The data represented by the lines labelled B-E correspond to the bar graphs depicted in Fig. 2B-E. **B-E.** Bar graphs showing military and R&D expenditures for discrete comparisons between countries. B is an example of a concordant pair. C is an example of a discordant pair. D is an example of a near-tie for R&D expenditure only. E is an example of a near-tie for military expenditure only.

Aside from these discrete comparisons, we can also examine the global distribution of coordinates along the Cartesian plane. Here, we note a skew in the number of coordinates located in Quadrant III (Fig. 2A). As illustrated and discussed above (Fig. 2C-D), these coordinates correspond to discordant pairs in which one country allocated more of their budget to military expenditures while the other allocated more to R&D, hence these points are located below the x-axis. Because these coordinates are in Quadrant III (where $\Delta military > \Delta R\&D$), we can further surmise that the difference in military spending between these countries was greater than the difference in spending on R&D.

Taken together, we can see that there is little association between the rank order of military vs. R&D expenditure between countries (τ=-0.02) insofar as there are approximately equal numbers of concordant and discordant pairs; however, there are meaningful differences between discrete comparisons that are reflected in the global distribution of coordinates that highlight a certain degree of heterogeneity in the underlying data. Through this graphical representation of Kendall's τ, these features can be discretized and spatially organized along the graph such that it becomes easier to visualize and identify the differences.

*Alternative Graphing Styles*

Thus far, we have relied on lines which connect coordinates corresponding to our observations to graph Kendall's τ. This method works well for smaller data sets (Fig. 1B, 2A; Supplementary File 1); however, as the number of observations increases, the graph becomes more cluttered and difficult to interpret (Fig. 3A; Supplementary File 1). Therefore, we propose two alternative graphing styles to remedy this issue. First, density contour plots can be used to represent the distribution and density of observations along the graph (Fig. 3B). Alternatively, vectors can be used to illustrate key features of Kendall's τ using what we have termed a "clock plot" due to its similar appearance to a clock (Fig. 3C). Here, the mean of a set of vectors corresponding to all concordant pairs is calculated and then represented by a black line which—as a result of being concordant—will always be located above the x-axis. The mean of all discordant vector pairs is represented by a black line which—as a result of being discordant—will always be located below the x-axis. Finally, the mean of all concordant and discordant vector pairs is represented by a red line—the position of which depends on and directly reflects the τ value (Fig. 3C). In other words, one can quickly surmise that the τ value is below 0 if the red summary vector is below the x-axis, above 0 if the vector is above the x-axis, and exactly 0 if the vector is parallel to the x-axis. Importantly, the various methods described here can be combined to provide a variety of different graphing options (Fig. 4).

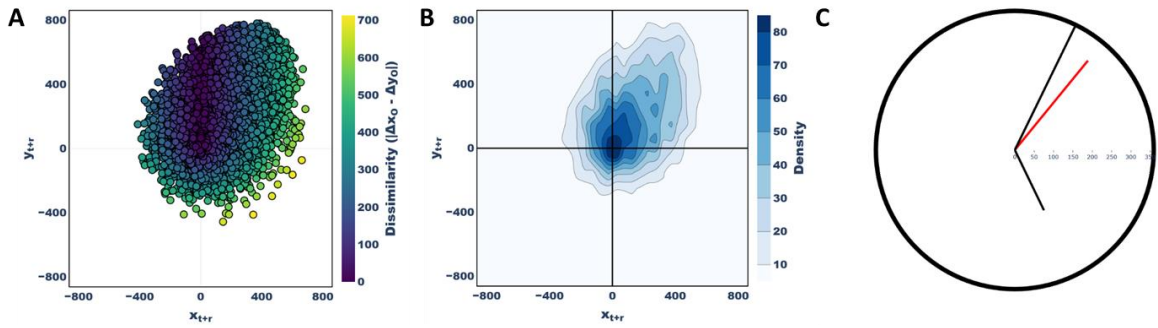

Figure 3. Alternative graphing styles. **A.** A visual representation of Kendall's τ using a large data set which is difficult to interpret because of the high degree of overlap between points. **B.** An alternative representation of the large data set shown in Fig. 3A using a density plot. **C.** An alternative representation of the large data set shown in Fig. 3A using a "clock plot". The mean vector of all concordant pairs is represented by a black line located above the x-axis. The mean vector of all discordant pairs is represented by a black line located below the x-axis. The mean vector of all concordant and discordant pairs, which corresponds directly to the τ value, is represented by a red line which may be located either above or below the x-axis. Here, the red line is located above the x-axis indicating that τ > 0.

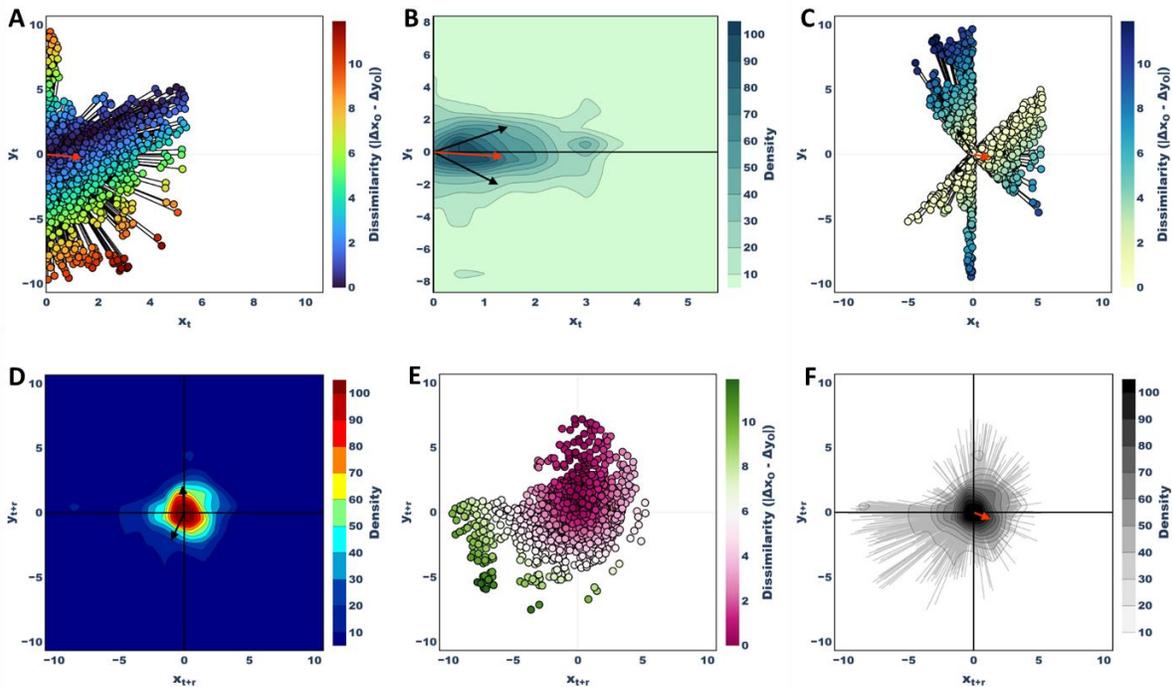

Figure 4. Combining graphing styles. Examples of how different graphing styles can be combined to produce different aesthetics, including: **A.** Translation, lines connecting coordinates, and a summary vector; **B.** Translation, a density plot, the mean concordant vector, the mean discordant vector, and the summary vector; **C.** Translation, lines connecting coordinates, the mean concordant vector, the mean discordant vector, and the summary vector; **D.** Translation and rotation, a density plot, the mean concordant vector, and the mean discordant vector; **E.** Translation and rotation with coordinates alone; **F.** Translation and rotation, lines connecting coordinates, a density plot, and the summary vector.

### *Representative τ Values*

In order to illustrate how graphical patterns will change with respect to different τ values, we provide a hypothetical set of figures derived from data corresponding to τ values ranging from 0.911 to -0.911 for several of the graphing styles described above (Fig. 5, Supplementary File 1). Here, we observe distinct and recognizable patterns that correspond to different τ values. Namely, as the τ value approaches 1, the proportion of coordinates terminating in the upper quadrants increases. As the τ value approaches -1, the proportion of coordinates terminating in the lower quadrants increases. τ values in between 1 and -1 will result in coordinates that become gradually more dispersed. Indeed, as the τ value approaches 0, the number of coordinates terminating in the upper and lower quadrants will be equal, corresponding to an equal number of concordant and discordant pairs.

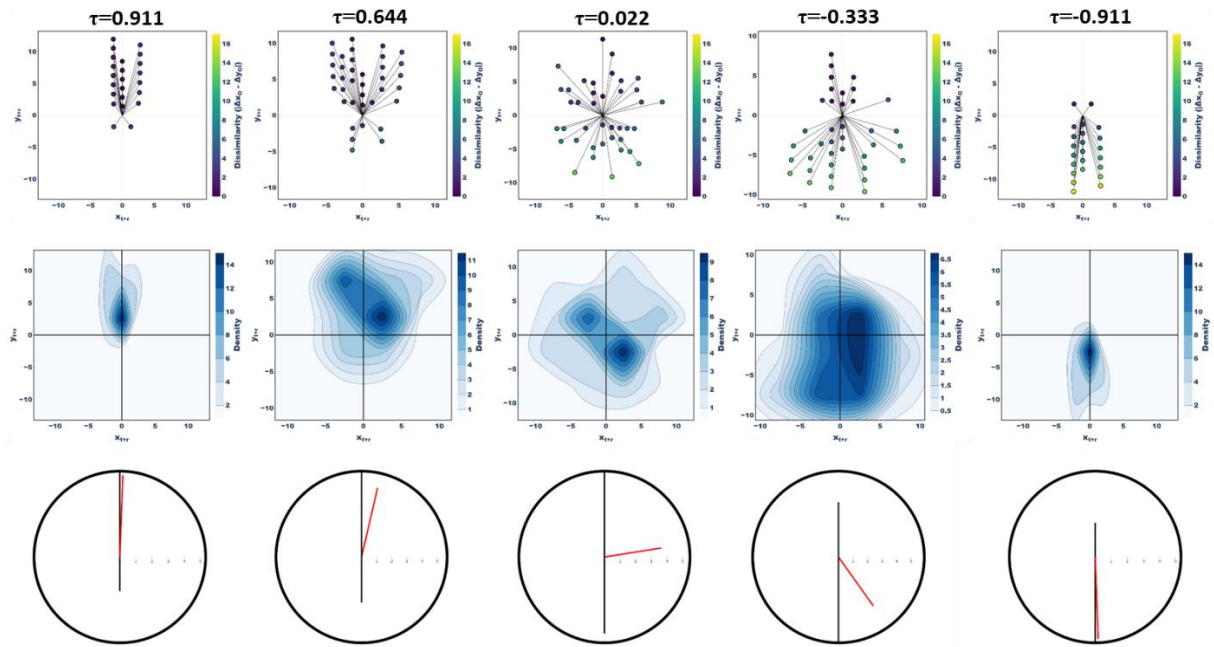

Figure 5. Graphical patterns that emerge from different τ values. Rows depict different graphing styles, including line graphs, density plots, and clock plots, respectively. Columns correspond to τ values ranging from 0.911 to -0.911, respectively.

## *The Same τ Value can Produce Fundamentally Different Graphs*

Another interesting feature that is not readily apparent from the τ statistic alone but that emerges from the graphing methods outlined in this report is that a single τ value can result in profoundly different graphical representations (Fig. 6A-F, Supplementary File 1). This is because different rank orders can alter the relationship between observations while maintaining the same overall τ value. We therefore suggest that graphing Kendall's τ can be a powerful method for obtaining a deeper understanding of the meaningful relationships that exist within the underlying data, especially in situations where the identity of the ranked data sets (e.g., country A vs. country B) or their components (e.g., military expenditure vs. R&D expenditure) are of particular interest.

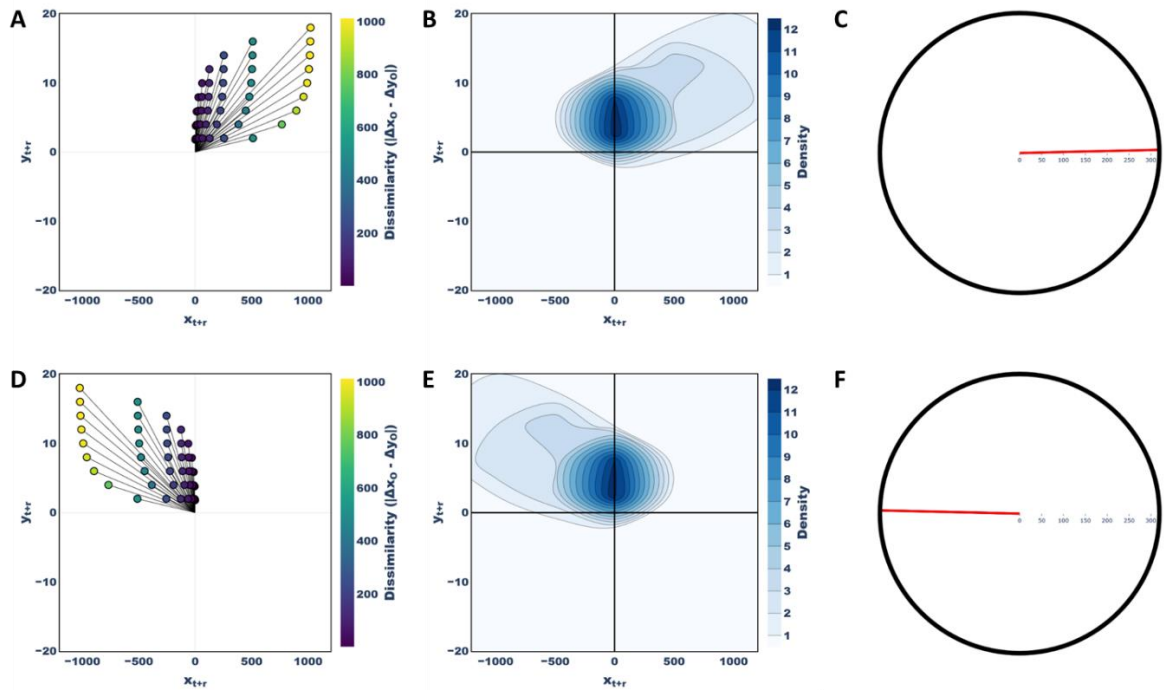

Figure 6. The same τ value can produce fundamentally different graphs. **A-C.** Representative data (τ=1) where $\Delta x_0$ is consistently greater than $\Delta y_0$. **D-F.** Representative data (τ=1) where $\Delta y_0$ is consistently greater than $\Delta x_0$.

## *Interactive App for Visualizing Kendall's τ*

To facilitate the visualization of Kendall's τ, we provide a Python-based web application script using Plotly's Dash library (Plotly Technologies Inc., Montreal, QC, Canada) that enables users to upload ranked data and generate informative graphs using the novel graphing techniques described in this manuscript, either in combination or alone (Supplementary File 3). One of the distinctive features of this app lies in its provision of diverse customization options, allowing users to customize the appearance of the figure to best suit their specific needs and preferences. In doing so, we hope to enhance the accessibility of our graphing technique and facilitate the analysis and communication of ranked data relationships.

*Conclusions*

Ranked data is a fundamental data type used in research across many fields of study for which a variety statistical tests have been developed (Friedman 1937, 1939; Kendall 1938; Milton 1940; Spearman 1987, Wilcoxon 1945; Pirttilä and Uusitalo 2010; Weber and Titman 2019, Ferguson et al. 2023, Fincham et al. 2023). Here, we examine Kendall's τ coefficient, a statistical tool used for analyzing ranked data by measuring the strength of the association between two ordinal or continuous variables (Kendall 1938). However, nearly a century after its initial conception, there remain few methods for graphing Kendall's τ (Holmes 1928, Davis and Chen 2007). In order to address this gap and provide researchers with additional options for visualizing their data, we provide an approach that leverages a series of rigid Euclidean transformations to map rank pairs onto discretely meaningful quadrants along a Cartesian plane. Our method yields a graphical representation of rank correlation that captures both the proportion of concordant and discordant pairs and highlights heterogeneity in the relationships between discrete pairs of observations. In doing so, we hope to provide the scientific community with additional graphing options to facilitate the visualization and rapid interpretation of ranked data.

**Acknowledgements:** We thank Dr. L.J. Zwiebel of Vanderbilt University for his generous support throughout the course of this work. We also thank Dr. A. Leich Hilbun for her thoughtful comments and suggestions on this manuscript.

**Funding:** This work was supported by a grant from the National Institutes of Health (NIGMS/RO1GM128336) to L.J.Z. and with endowment funding from Vanderbilt University.

**Disclosure Statement:** The authors declare no competing or financial interests.

**Availability of Data and Materials:** All data generated or analyzed are included in this published article and its supplementary information files. The Python-based web application script is available in Supplementary File 3 and has been made publicly available for download on Figshare repository at doi.org/10.6084/m9.figshare.24060996.

*PRE-PRINT*

**Figure Legends**

Figure 1. Method for visualizing Kendall's τ using rigid Euclidean transformations. **A.** Representative ranked data plotted as coordinates on a Cartesian plane. Each coordinate represents an item which has been ranked by two entities—$x_0$ or $y_0$—and the lines between coordinates signify the pairwise comparisons made during the evaluation of Kendall's τ. **B.** The result of translating pairs of observations to the origin and then rotating the lines by doubling their angle (2*θ); note that the axes labels have been updated to denote this transformation ($x_{t+r}$ or $y_{t+r}$). The area above the x-axis in white denotes concordant pairs, and the area below the x-axis in gray represents discordant pairs. **C.** A heatmap demonstrating how dissimilarity within and between rankings, as measured by the absolute difference between $\Delta x_0$ or $\Delta y_0$, changes with respect to location along the Cartesian plane.

Figure 2. A representative, real-world example using governments' expenditures on military vs. R&D. **A.** A visual representation of Kendall's τ using military and R&D expenditure (as a percentage of GDP) from 62 different countries (Supplementary File 1) (World Bank, N.d.a, N.d.b). The data represented by the lines labelled B-E correspond to the bar graphs depicted in Fig. 2B-E. **B-E.** Bar graphs showing military and R&D expenditures for discrete comparisons between countries. B is an example of a concordant pair. C is an example of a discordant pair. D is an example of a near-tie for R&D expenditure only. E is an example of a near-tie for military expenditure only.

Figure 3. Alternative graphing styles. **A.** A visual representation of Kendall's τ using a large data set which is difficult to interpret because of the high degree of overlap between points. **B.** An alternative representation of the large data set shown in Fig. 3A using a density plot. **C.** An alternative representation of the large data set shown in Fig. 3A using a "clock plot". The mean vector of all concordant pairs is represented by a black line located above the x-axis. The mean vector of all discordant pairs is represented by a black line located below the x-axis. The mean vector of all concordant and discordant pairs, which corresponds directly to the τ value, is represented by a red line which may be located either above or below the x-axis. Here, the red line is located above the x-axis indicating that τ > 0.

Figure 4. Combining graphing styles. Examples of how different graphing styles can be combined to produce different aesthetics, including: **A.** Translation, lines connecting coordinates, and a summary vector; **B.** Translation, a density plot, the mean concordant vector, the mean discordant vector, and the summary vector; **C.** Translation, lines connecting coordinates, the mean concordant vector, the mean discordant vector, and the summary vector; **D.** Translation and rotation, a density plot, the mean concordant vector, and the mean discordant vector; **E.** Translation and rotation with coordinates alone; **F.** Translation and rotation, lines connecting coordinates, a density plot, and the summary vector.

Figure 5. Graphical patterns that emerge from different $\tau$ values. Rows depict different graphing styles, including line graphs, density plots, and clock plots, respectively. Columns correspond to $\tau$ values ranging from 0.911 to -0.911, respectively.

Figure 6. The same $\tau$ value can produce fundamentally different graphs. **A-C.** Representative data ($\tau=1$) where $\Delta x_0$ is consistently greater than $\Delta y_0$. **D-F.** Representative data ($\tau=1$) where $\Delta y_0$ is consistently greater than $\Delta x_0$.

Figure S1. Alternative graphing styles using translation alone. **A-B.** Coordinate pairs which have been translated to the origin. The decision regarding which coordinate to translate to the origin is arbitrary because the quadrants which contain information about concordance/discordance and the dissimilarity between ranks are mirrored and reflected. **C-D.** Heatmaps demonstrating how dissimilarity within and between rankings, as measured by the absolute difference between $\Delta x_0$ or $\Delta y_0$, changes with respect to location along the Cartesian plane for the figures shown in A and B, respectively.

Supplementary File 1. Raw data.

Supplementary File 2. A demonstration of the rigid transformations used to visualize Kendall's $\tau$.

Supplementary File 3. Interactive Python-based script for graphing Kendall's $\tau$ using Plotly's Dash library (Plotly Technologies Inc., Montreal, QC, Canada).